\documentclass[12pt,pre,aps,showpacs,preprint]{revtex4}

\usepackage{graphics}
\usepackage{graphicx}
\usepackage{amsfonts}
\usepackage{textcomp}
\usepackage{amssymb}
\usepackage{amsmath}

\begin{document}

\title{Exact Constructions of a Family of Dense Periodic Packings of Tetrahedra}

\author{S. Torquato$^{1,2,3,4,5,6}$ and Y. Jiao$^5$}


\affiliation{$^1$Department of Chemistry, Princeton University,
Princeton New Jersey 08544, USA}




\affiliation{$^2$Department of Physics, Princeton University,
Princeton New Jersey 08544, USA}

\affiliation{$^3$Princeton Center for Theoretical Science,
Princeton University, Princeton New Jersey 08544, USA}

\affiliation{$^4$Program in Applied and Computational Mathematics,
Princeton University, Princeton New Jersey 08544, USA}

\affiliation{$^5$School of Natural Sciences, Institute for
Advanced Study, Princeton NJ 08540}

\affiliation{$^6$Department of Mechanical and Aerospace
Engineering, Princeton University, Princeton New Jersey 08544,
USA}


\begin{abstract}
The determination of the densest packings of regular tetrahedra
(one of the five Platonic solids) is attracting great attention as
evidenced by the rapid pace at which packing records are being
broken and the fascinating packing structures that have emerged.
Here we provide the most  general analytical formulation to date
to construct dense periodic packings of tetrahedra with four
particles per fundamental cell. This  analysis results in
six-parameter family of dense tetrahedron packings that includes
as special cases recently discovered ``dimer" packings of
tetrahedra, including the densest known packings with density
$\phi= \frac{4000}{4671} = 0.856347\ldots$. This study strongly
suggests that the latter set of packings are the densest among all
packings with a four-particle basis. Whether they are the densest
packings of tetrahedra among all packings is an open question, but
we offer remarks about this issue. Moreover, we describe a
procedure that provides estimates of upper bounds on the maximal
density of tetrahedron packings, which could aid in assessing the
packing efficiency of candidate dense packings.
\end{abstract}


\maketitle

\section{Introduction}

Dense packings of nonoverlapping solid objects or particles are
ubiquitous in synthetic and natural situations. Packing problems
arise in technological contexts, such as the packaging industries,
agriculture (e.g., grains in silos), and solid-rocket propellants,
and underlie the structure of a multitude of biological systems
(e.g., tissue structure, cell membranes, and phyllotaxis). Dense
particle packings are intimately related to the structure of
low-temperature states of condensed matter, such as liquids,
glasses and crystals \cite{We71,Za83,Chaik95,To02a}. In the last
decade, scientific attention has broadened from the study of dense
packings of spheres (the simplest shape that does not tile
Euclidean space)
\cite{Ga31,Be60,Co93,Li98,To00b,Oh02,Co03,Ha05,To06b,Co09} to
dense packings of disordered  \cite{Do04a,Do07,Ma09,Ze09,Ji10} and
ordered \cite{Be00,Do04d,Ji09a} nonspherical particles. In
addition, the equilibrium phase behavior and transport properties
of hard nonspherical particles have been of topics of great
interest  \cite{Allen93, Frenkel83, Frenkel84, Schw08}.
Nonsphericity introduces rotational degrees of freedom not present
in sphere packings, and can dramatically alter the characteristics
from those of sphere packings.

A basic characteristic of a packing in $d$-dimensional Euclidean
space $\mathbb{R}^d$ is its density $\phi$, defined to be the
fraction of $\mathbb{R}^d$ that is covered by the particles.
A problem that has been of great scientific interest for centuries
is the determination of the densest arrangement(s) of particles
that do not tile space and the associated maximal density
$\phi_{max}$. For generally shaped particles, finding the densest
packings is notoriously difficult. This salient point is
summarized well by Henry Cohn \cite{Co09b} who recently remarked,
``For most grain shapes we cannot guess or even closely
approximate the answer, let alone prove it, and it is difficult to
develop even a qualitative understanding of the effects of grain
shape on packing density.'' Until recently, very little was known
about the densest packings of polyhedral particles. The difficulty
in obtaining dense packings of polyhedra is related to their
complex rotational degrees of freedom and to the non-smooth nature
of their shapes \cite{To09b,To09c}.

Recently, we set out to attempt to determine the densest known
packings of the Platonic and Archimedean solids
\cite{To09b,To09c}. It was shown that the central symmetry of the
majority of the Platonic and Archimedean solids distinguish their
dense packing arrangements from those of the non-centrally
symmetric ones in a fundamental way. (A particle is centrally
symmetric if it has a center $C$ that bisects every chord through
$C$ connecting any two boundary points of the particle; i.e., the
center is a point of inversion symmetry.) The tetrahedron is  the
only Platonic solid that lacks central symmetry, an attribute that
geometrically frustrates it to a greater degree than the majority
of the remaining solids in this set that do not tile space
\cite{To09c}. A number of organizing principles emerged in the
form of conjectures for polyhedra as well as other nonspherical
shapes. In the case of polyhedra, the following three are
particularly relevant and will be applied later in the paper to
remark on the optimality of tetrahedron packings:

\begin{itemize}
\itemsep 0.05in \item {\bf Conjecture 1}: {\sl The densest
packings of the centrally symmetric Platonic and Archimedean
solids are given by their corresponding optimal Bravais lattice
packings.}

\item {\bf Conjecture 2}: {\sl The densest packing of any convex,
congruent polyhedron without central symmetry generally is not a
Bravais lattice packing, i.e., the set of such polyhedra whose
optimal packing is not a Bravais lattice is overwhelmingly larger
than the set whose optimal packing is a Bravais lattice. }

\item {\bf Conjecture 3}: {\sl The densest packings of congruent,
centrally symmetric particles that do not possesses three
equivalent principle axes (e.g., ellipsoids) generally cannot be
Bravais lattices.}
\end{itemize}

Conjecture 1 is the analog of Kepler's sphere conjecture for the
centrally symmetric Platonic and Archimedean solids. In this
sense, such  solids behave similarly to spheres in that their
densest packings are lattice arrangements and (except for the cube
and the truncated octahedron) are geometrically frustrated like
spheres. Conjecture 2 has been shown by Conway and Torquato
\cite{Co06} to be true for both the tetrahedron and Archimedean
non-centrally symmetric truncated tetrahedron, the latter of which
can be arranged in a ``uniform" non-Bravais lattice packing with
density at least as high as $\frac{23}{24}=0.958333\ldots$. (A
\textit{uniform} packing has a symmetry operation, in this case
the point inversion symmetry, that takes any particle into
another.) Although Conjecture 3 is explicitly stated in its
current form here for the first time, it was strongly implied in
Ref.~\cite{To09c}.

It was Conway and Torquato's
investigation \cite{Co06} that has spurred the flurry of activity in the last
several years to find the densest packings of tetrahedra. There
have been many twists and unexpected turns since 2006 that have
led to the dense packings of tetrahedra that we report here. Therefore, to place our present
results in their proper context, it is instructive to review
briefly the developments since 2006.

\begin{table}
\centering \caption{A brief summary of the dense non-lattice
packings of  tetrahedra. The name of the packing is given along
with the year that it was discovered. Here  $\phi$ is the packing
density and $N$ is the number of tetrahedra per fundamental cell.}
\begin{tabular}{c@{\hspace{0.45cm}}c@{\hspace{0.45cm}}c@{\hspace{0.45cm}}c}
\hline\hline
Packing & Year & $\phi$ &  $N$  \\
\hline
Uniform I  \cite{Co06}  & 2006   & $\frac{2}{3}=0.666666\ldots$   & 2 \\
Welsh     \cite{Co06}      & 2006   & $\frac{17}{24}=0.708333\ldots$   & 34  \\
Icosahedral   \cite{Co06}  & 2006   & $0.716559\ldots$   & 20  \\
Uniform II \cite{Ka09} & 2009 & $\frac{139+40\sqrt10}{369}=0.719488\ldots$ & 2 \\
Wagon Wheels   \cite{Ch08} & 2008   & $0.778615\ldots$   & 18  \\
Improved Wagon Wheels \cite{To09b} & 2009 & $0.782021\ldots$ & 72 \\
Disordered Wagon Wheels \cite{To09c}& 2009 & $0.822637\ldots$ & 314 \\
Ring Stacks \cite{Gl09} &2009& $0.8503\ldots$ & 82 \\
Uniform III \cite{Ka09} & 2009 &$\frac{100}{117}=0.854700\ldots$ & 4\\
Dimer-Uniform I \cite{To10}& 2009 & $\frac{12250}{14319} = 0.855506\ldots$& 4\\
Dimer-Uniform II \cite{Chen10}& 2010 & $\frac{4000}{4671} = 0.856347\ldots$& 4\\
\hline\hline
\end{tabular}
\label{tab1}
\end{table}

First, we note that the densest Bravais-lattice packing of
tetrahedra (requiring one tetrahedron per fundamental cell such
that each tetrahedron in the packing has the same orientation as
the others) has a packing fraction $\phi =\frac{18}{49}=0.367\ldots$ and
each tetrahedron touches 14 others \cite{Ho70}. Conway and
Torquato \cite{Co06} showed that the densest packings of
tetrahedra cannot be Bravais lattices by analytically constructing
several such packings with densities that are substantially larger
than $\frac{18}{49}$. (A non-Bravais lattice packing contains multiple
particles, with generally different orientations, per fundamental
cell, which is periodically replicated in $\mathbb{R}^d$.) One
such packing is a ``uniform" packing with density $\phi=\frac{2}{3}$ and
two particles per fundamental cell. The so-called ``Welsh" packing
has a density $\phi=0.708333\ldots$ and 34 particles per
fundamental cell. Yet another non-Bravais lattice packing with
density  $\phi = 0.716559\ldots$ is based on the filling of
``imaginary" icosahedra with the densest arrangement of 20
tetrahedra and then arranging the imaginary icosahedra in their
densest lattice packing configuration. The densities of both the
Welsh and Icosahedral packings can be further improved by certain
particle displacements  \cite{Co06}. Using imperfect
``tetrahedral" dice, Chaikin {et al.} \cite{Ch07} experimentally
generated jammed disordered packings with $\phi \approx 0.75$.
Employing physical models and a computer algebra system, Chen
\cite{Ch08} discovered a remarkably dense periodic arrangement of
tetrahedra with $\phi=0.7786\ldots$, which exceeds the density
($\phi_{max}=\pi/\sqrt{18} =0.7404\ldots$) of the densest sphere
packing by an appreciable amount. We have called this the
``wagon-wheels" packing \cite{To09b,To09c}.

Torquato and Jiao \cite{To09b} devised and applied an optimization
scheme, called  the adaptive-shrinking-cell (ASC) method, that
used an initial configuration based on the wagon-wheels packing to
yield a non-Bravais lattice packing consisting of 72 tetrahedra
per fundamental cell with a density $\phi =0.782\ldots$
\cite{To09b}. Using 314 particles per fundamental cell and
starting from an ``equilibrated" low-density liquid configuration,
the same authors were able to improve the density to $\phi
=0.823\ldots$ \cite{To09c}. This packing arrangement interestingly
lacks long-range order. Haji-Akbari {\it et al.} \cite{Gl09}
numerically constructed a periodic packing of tetrahedra made of
parallel stacks of ``rings" around ``pentagonal'' dipyramids
consisting of 82 particles per fundamental cell and a density
$\phi=0.8503\ldots$. More recently, Kallus {\it et al.}
\cite{Ka09} found a remarkably simple uniform packing of
tetrahedra with high symmetry consisting of only four particles
per fundamental cell with density
$\phi=\frac{100}{117}=0.854700\ldots$. We subsequently
presented an analytical formulation to construct dense uniform
dimer packings of tetrahedra and employed it to obtain a three-parameter family of
packings. (A dimer is composed of a pair of regular tetrahedra that exactly share a common face.
A uniform dimer packing of tetrahedra takes any dimer via a point-inversion symmetry
operation into another.)  Making an assumption about one of these parameters
resulted in a two-parameter family, including those with density as high as
$\phi=\frac{12250}{14319}=0.855506\ldots$ \cite{To10}.
Chen {\it et al.} \cite{Chen10} recognized that we made such an assumption
and employed a similar formalism
to obtain a three-parameter family of tetrahedron packings,
including the densest known dimer packings of tetrahedra with
a density $\phi = \frac{4000}{4671} = 0.856347\ldots$. Table
\ref{tab1} summarizes some of the packing characteristics of the
non-Bravais lattice packings of tetrahedra.

In the following section, we provide the details of our more general formulation
and construct a six-parameter family of dense tetrahedron packings.
We will show that our formalism includes as special cases all of the recently
discovered four-particle basis packings \cite{Ka09,To10,Chen10}.
Our analysis strongly
suggests that the optima among this  set of packings provide the densest arrangements among
all packings with a four-particle basis.
In Sec.~\ref{bounds}, we describe a procedure that provides estimates
of upper bounds on the maximal density of tetrahedron
packings, which could aid in assessing the packing efficiency of
candidate dense packings. In Sec.~\ref{discuss},
we make concluding remarks, including comments  on the optimality of the densest known
dimer packings of tetrahedra.

\section{Analytical Constructions of Dense Packings of Tetrahedra}

Inspired by the work of Kallus et al. \cite{Ka09}, we have applied
the adaptive-shrinking-cell (ASC) optimization scheme to  examine
comprehensively packings with a considerably small number of
particles per fundamental cell (from 2 to 32) than we  have used
in the past \cite{To09b, To09c}. The ASC scheme employs both a
sequential search of the configurational space of the particles
and the space of lattices via an adaptive fundamental cell that
deforms and shrinks on average to obtain dense packings. A dense
packing with 8-particle basis that emerged from this numerical
investigation suggested that it was composed of two very similar
fundamental cells, each containing 4 particles. Using one of the
4-particle basis configurations, we were able to find packings
with density $\phi = 0.8551034\ldots$ that exceeded the highest
density packings with $\phi=\frac{100}{117}=0.854708\ldots$
constructed by Kallus {\it et al.} Even though our packings
possess a type of point inversion symmetry, they are not as
symmetric as the densest packings reported in Ref. \cite{Ka09}, as
we now explain.

The four tetrahedra in the fundamental cell in our dense numerically
generated packings formed two contacting ``dimers''. A dimer is
composed of a pair of regular tetrahedra with unit edge length
that exactly share a common face. The compound object consisting
of the two contacting dimers possesses point inversion symmetry,
with the inversion center at the centroid of the contacting region
on the faces. A Bravais lattice possesses point inversion symmetry
about the lattice points and the centroids of the fundamental
cells. By placing the symmetry center of the two-dimer compound on
the centroids (or the lattice points), we construct packings that
generally possess point inversion symmetry only about the symmetry
centers of the two-dimer compound.  Besides the centroids of the
fundamental cell, all of the half-integer lattice points are also
inversion symmetry centers of the packing. We call such structures
{\it dimer-uniform} packings, since the inversion symmetry acts to
take any {\it dimer} into another. Such packings should be
distinguished from the more symmetric {\it uniform} (or
transitive) packings of tetrahedra in which the symmetry operation
acts to take any {\it tetrahedron} into another, such as the ones
found in Refs.~\cite{Co06} and ~\cite{Ka09} (see
Table~\ref{tab1}). The latter have almost as much  symmetry as a
Bravais lattice, except that the centroids of the particles are
not just characterized by simple translational symmetry.

\subsection{General Formalism: Six-Parameter Family of Dense Tetrahedron Packings}

We now describe our general analytical formulation to construct
dense tetrahedron packings by relaxing the symmetry conditions on
the contacting dimers in detail. In particular, we orient the
3-fold rotational symmetry axis of one of the dimers in an
arbitrary direction (say the $z$-direction of a Cartesian
coordinate system), and then fix the origin of the lattice vectors
at the centroid of this dimer. (The centroid is located at the
center of the contacting faces of the two tetrahedra that comprise
the dimer.) Then we place the second dimer in contact with the
first one such that there is a center of inversion symmetry that
takes one dimer to the other, which implies a face-to-face contact
between the two dimers.

The problem of determining the analytical constructions then
amounts to determining 12 equations for the 12 unknowns. Nine of
the 12 unknowns arise from the three unknown lattice vectors, each
of which contains three unknown components. The other 3 unknowns
derive from the components of the centroid of the second dimer.


In particular, we let the centroid of the dimer at the origin be
denoted by ${\bf c}_0 = (0, 0, 0)$, and the centroid of the other
dimer be denoted by ${\bf c}_1 = (\eta_1, \eta_2, \eta_3)$. The
vertices of the two dimers associated with ${\bf c}_0$ and ${\bf
c}_1$ are given by ${\bf v}_A = (\frac{1}{2},
\frac{1}{2\sqrt3},0)$, ${\bf v}_B = (-\frac{1}{2},
\frac{1}{2\sqrt3},0)$, ${\bf v}_C = (0, -\frac{1}{\sqrt3},0)$,
${\bf v}_D = (0, 0, \sqrt{\frac{2}{3}})$, ${\bf v}_E = (0, 0,
-\sqrt{\frac{2}{3}})$ and ${\bf v}^*_A = -{\bf v}_A+{\bf c}_1$,
${\bf v}^*_B = -{\bf v}_B+{\bf c}_1$, ${\bf v}^*_C = -{\bf
v}_C+{\bf c}_1$, ${\bf v}^*_D = -{\bf v}_D+{\bf c}_1$, ${\bf
v}^*_E = -{\bf v}_E+{\bf c}_1$, respectively. In addition, let the
lattice vectors be ${\boldsymbol{\lambda}}_1 = (-\alpha_1,
-\alpha_2, -\alpha_3)$, ${\boldsymbol{\lambda}}_2 = (\beta_1,
-\beta_2, -\beta_3)$, and ${\boldsymbol{\lambda}}_3 = (\gamma_1,
\gamma_2, \gamma_3)$. The 12 components of the four vectors ${\bf
c}_1$ and ${\boldsymbol{\lambda}}_i$, ($i~=~1,2,3$) are the
variables that determine the packing. Note in the above general
set-up, we assume no particular symmetry of the packings.

\begin{figure}[ht]
\begin{center}
$\begin{array}{c}
\includegraphics[height=8.5cm, keepaspectratio]{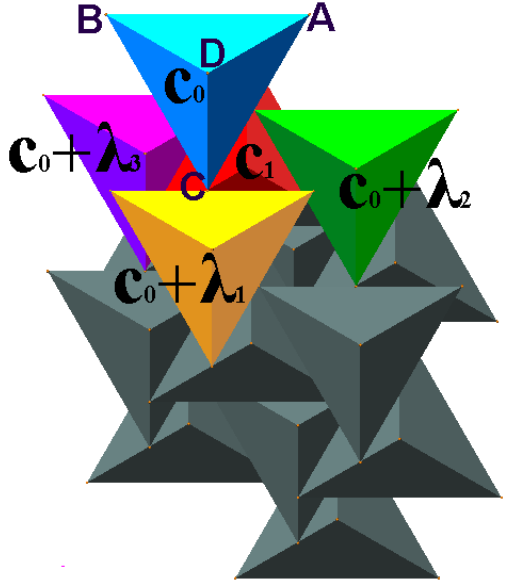} \\
\end{array}$
\end{center}
\caption{(color online). The geometrical set up for the general formulation of the
analytical constructions of dimer packings of tetrahedra. The
dimer (two tetrahedra sharing a common face) centered at ${\bf
c}_0$ is shown in blue with the vertices labeled (note that ${\bf
v}_E$ whose projection coincides with ${\bf v}_D$ is not shown).
The dimer centered at ${\bf c}_1$ is shown in red. The dimer
centered at $({\bf c}_0+\boldsymbol{\lambda}_1)$ is shown in
yellow. The dimer centered at $({\bf c}_0+\boldsymbol{\lambda}_2)$
is shown in green. The dimer centered at $({\bf
c}_0+\boldsymbol{\lambda}_3)$ is shown in purple. Observe that
in the perspective shown the dimers appear as if they were single
tetrahedra.} \label{fig0}
\end{figure}

In our packings, the dimers have 8 possible face-to-face contacts,
6 possible edge-to-edge contacts and 4 possible vertex-to-face contacts.
Each contact provides a condition that either reduces the number of variables
for the packing or constrains their feasible values.


A face-to-face contact requires that the projection of the vector
distance between the centroids of the two dimers on the contacting
face normal is equal to the diameter of the insphere (the largest possible
sphere that can be inscribed into a tetrahedron) of the tetrahedra.
A nonoverlapping condition associated with a face-to-face
contact requires that the projection of the vector
distance between the centroids of the two dimers on the contacting
face normal is greater than the diameter of the insphere.
The 8 possible face-to-face contacts
are between the dimer pairs with the centroids at $\{{\bf c}_1,
{\bf c}_0\}$, $\{{\bf c}_1, {\bf c}_0+{\boldsymbol{\lambda}}_1\}$, $\{{\bf c}_1,
{\bf c}_0+{\boldsymbol{\lambda}}_2\}$, $\{{\bf
c}_0+\boldsymbol{\lambda}_3, {\bf c}_1\}$, $\{{\bf
c}_0+\boldsymbol{\lambda}_3, {\bf c}_1-\boldsymbol{\lambda}_2\}$,
$\{{\bf c}_0+\boldsymbol{\lambda}_3, {\bf
c}_1+\boldsymbol{\lambda}_1-\boldsymbol{\lambda}_2\}$, $\{{\bf
c}_0, {\bf c}_1-\boldsymbol{\lambda}_1+\boldsymbol{\lambda}_3\}$,
and $\{{\bf c}_0, {\bf
c}_1+\boldsymbol{\lambda}_1-\boldsymbol{\lambda}_2-2\boldsymbol{\lambda}_3\}$.
The possible contact between dimer pairs at $\{{\bf u}_i, {\bf u}_j\}$
requires
\begin{equation}
\label{eq001} ({\bf u}_i-{\bf u}_j)\cdot {\bf n}_{ij} \ge
\frac{2\sqrt6}{9},
\end{equation}
where ${\bf n}_{ij}$ is unit outward contacting face normal of the
dimer at ${\bf u}_j$. In the following discussion, we explicitly
provide these equations.

First, we note that ${\bf c}_1$ can be completely determined by
considering the contacts that the dimer associated with it makes
with three other dimers centered at ${\bf c}_0$, $({\bf
c}_0+{\boldsymbol{\lambda}}_1)$ and $({\bf
c}_0+{\boldsymbol{\lambda}}_2)$ (see Fig.~\ref{fig0}). Note that
$\boldsymbol{\lambda}_1$ and $\boldsymbol{\lambda}_2$ are two of
the lattice vectors. The equations associated with these three
contacts are respectively given by
\begin{equation}
\label{eq5001}
\begin{array}{c}
\displaystyle{\frac{1}{6} \eta_1 - \frac{1}{6\sqrt3} \eta_2 -
\frac{\sqrt2}{12\sqrt3}\eta_3 = \frac{1}{9},} \\
\end{array}
\end{equation}

\begin{equation}
\label{eq5002}
\begin{array}{c}
 \displaystyle{\frac{1}{3\sqrt3}
(\eta_2+\alpha_2) -\frac{\sqrt2}{12\sqrt3} (\eta_3+\alpha_3) =
\frac{1}{9},}\\
\end{array}
\end{equation}

\begin{equation}
\label{eq5003}
\begin{array}{c}
\displaystyle{-\frac{1}{6}(\eta_1-\beta_1)-\frac{1}{6\sqrt3}(\eta_2+\beta_2)-\frac{\sqrt2}{12\sqrt3}(\eta_3+\beta_3)
= \frac{1}{9}.}
\end{array}
\end{equation}

\noindent By solving Eqs.~(\ref{eq5001})~-~(\ref{eq5003}), we can
obtain the components $(\eta_1, \eta_2, \eta_3)$ of ${\bf c}_1$,
i.e.,
\begin{equation}
\label{eq5004}
\begin{array}{c}
\displaystyle{\eta_1 =
\frac{1}{12}(6\beta_1-2\sqrt3\beta_2-\sqrt6\beta_3)},\\
\end{array}
\end{equation}

\begin{equation}
\label{eq5005}
\begin{array}{c}
\displaystyle{\eta_2 =
\frac{1}{12}(-8\alpha_2+2\sqrt2\alpha_3+2\sqrt3\beta_1-2\beta_2-\sqrt2\beta_3)},\\
\end{array}
\end{equation}

\begin{equation}
\label{eq5006}
\begin{array}{c}
\displaystyle{\eta_3 =
-\frac{1}{3\sqrt2}(4\sqrt3-4\alpha_2+\sqrt2\alpha_3-2\sqrt3\beta_1+2\beta_2+\sqrt2\beta_3)},\\
\end{array}
\end{equation}

Moreover, we note that ${\boldsymbol{\lambda}}_3$ can be
completely determined by considering the contacts between the
dimer centered at $({\bf c}_0+{\boldsymbol{\lambda}}_3)$ and at
${\bf c}_1$, $({\bf c}_1-{\boldsymbol{\lambda}}_2)$ and $({\bf
c}_1+{\boldsymbol{\lambda}}_1-{\boldsymbol{\lambda}}_2)$,
respectively. The equations associated with these three contacts
are given by
\begin{equation}
\label{eq5009}
\begin{array}{c}
\displaystyle{\frac{1}{6}(\eta_1-\gamma_1)
-\frac{1}{6\sqrt3}(\eta_2-\gamma_2)
+\frac{\sqrt2}{12\sqrt3}(\eta_3-\gamma_3) = \frac{1}{9}.}
\end{array}
\end{equation}

\begin{equation}
\label{eq5008}
\begin{array}{c}
 \displaystyle{\frac{1}{3\sqrt3}
(\eta_2+\beta_2-\gamma_2) +\frac{\sqrt2}{12\sqrt3}
(\eta_3+\beta_3-\gamma_3) =
\frac{1}{9},}\\
\end{array}
\end{equation}

\begin{equation}
\label{eq5007}
\begin{array}{c}
\displaystyle{-\frac{1}{6}(\eta_1+\alpha_1-\beta_1-\gamma_1)
-\frac{1}{6\sqrt3}(\eta_2+\alpha_2+\beta_2-\gamma_2)
+\frac{\sqrt2}{12\sqrt3}(\eta_3+\alpha_3+\beta_3-\gamma_3) =
\frac{1}{9}.}
\end{array}
\end{equation}

\noindent By solving Eqs.~(\ref{eq5009})~-~(\ref{eq5007}), we can
obtain the components $(\gamma_1, \gamma_2, \gamma_3)$ of
${\boldsymbol{\lambda}}_3$, i.e.,
\begin{equation}
\label{eq5010}
\begin{array}{c}
\displaystyle{\gamma_1 =
\frac{1}{12}(-6\alpha_1-2\sqrt3\alpha_2+\sqrt6 \alpha_3-2\sqrt6\beta_3)},\\
\end{array}
\end{equation}

\begin{equation}
\label{eq5011}
\begin{array}{c}
\displaystyle{\gamma_2 =
\frac{1}{12}(-2\sqrt3\alpha_1-10\alpha_2+3\sqrt2\alpha_3+8\beta_2)},\\
\end{array}
\end{equation}

\begin{equation}
\label{eq5012}
\begin{array}{c}
\displaystyle{\gamma_3 =
\frac{1}{3}(-4\sqrt6+\sqrt6\alpha_1+3\sqrt2\alpha_2-2\alpha_3+2\sqrt6\beta_1+\beta_3)}.\\
\end{array}
\end{equation}

By achieving the aforementioned 6 face-to-face contacts,
we can reduce the number of independent variables for the packing
from 12 to 6, with the remaining variables being $\alpha_i$
and $\beta_i$ ($i~=~1,2,3$). In other words, all of the lattice vectors and
the centroids of dimers are completely specified by the six parameters $\alpha_i$
and $\beta_i$, which gives the six-parameter family of dense packings, i.e.,

\begin{equation}
\label{eq50121}
\begin{array}{c}
\displaystyle{{\bf c}_1 = \left [{\begin{array}{c}\frac{1}{12}(6\beta_1-2\sqrt3\beta_2-\sqrt6\beta_3) \\
                                                   \frac{1}{12}(-8\alpha_2+2\sqrt2\alpha_3+2\sqrt3\beta_1-2\beta_2-\sqrt2 \beta_3) \\
                                                   -\frac{1}{3\sqrt2}(4\sqrt3-4\alpha_2+\sqrt2\alpha_3-2\sqrt3\beta_1+2\beta_2+\sqrt2\beta_3)
                                  \end{array}}\right]^T}, \\\\
\displaystyle{{\boldsymbol{\lambda}}_1 = (-\alpha_1,~ -\alpha_2,~ -\alpha_3)}, \quad\quad
\displaystyle{{\boldsymbol{\lambda}}_2 = (\beta_1,~-\beta_2,~-\beta_3)}, \\\\
\displaystyle{{\boldsymbol{\lambda}}_3 = \left [{\begin{array}{c}\frac{1}{12}(-6\alpha_1-2\sqrt3\alpha_2+\sqrt6\alpha_3-2\sqrt6\beta_3) \\
                                                   \frac{1}{12}(-2\sqrt3\alpha_1-10\alpha_2+3\sqrt2\alpha_3+8\beta_2) \\
                                                   \frac{1}{3}(-4\sqrt6+\sqrt6\alpha_1+3\sqrt2\alpha_2-2\alpha_3+2\sqrt6\beta_1+\beta_3)
                                                 \end{array}}\right]^T}.\\
\end{array}
\end{equation}
Note that these parameters
can not be varied completely independently of each other, i.e., they are related
by the additional nonoverlapping conditions which will be given in the ensuing
discussions.

There are two additional possible face-to-face
contacts between dimer pairs centered at $\{{\bf c}_0, {\bf
c}_1-{\boldsymbol{\lambda}}_1+{\boldsymbol{\lambda}}_3\}$, and
$\{{\bf c}_0, {\bf
c}_1+{\boldsymbol{\lambda}}_1-{\boldsymbol{\lambda}}_2-2{\boldsymbol{\lambda}}_3\}$.
The equations associated with these possible contacts are given by

\begin{equation}
\label{eq5013}
\begin{array}{c}
\displaystyle{-\frac{1}{6}(\eta_1+\alpha_1+\gamma_1)
-\frac{1}{6\sqrt3}(\eta_2+\alpha_2+\gamma_2)
-\frac{\sqrt2}{12\sqrt3}(\eta_3+\alpha_3+\gamma_3) \ge \frac{1}{9},}
\end{array}
\end{equation}

\begin{equation}
\label{eq5014}
\begin{array}{c}
\displaystyle{-\frac{1}{6}(\eta_1-\alpha_1-\beta_1-2\gamma_1)
+\frac{1}{6\sqrt3}(\eta_2-\alpha_2+\beta_2-2\gamma_2)
-\frac{\sqrt2}{12\sqrt3}(\eta_3-\alpha_3+\beta_3-2\gamma_3) \ge
\frac{1}{9}.}
\end{array}
\end{equation}

\noindent These conditions constrain the possible values of $\alpha_i$
and $\beta_i$.

An edge-to-edge contact requires that the projection of the
vector connecting the corresponding ends of two edges on the
common perpendicular line of the two edges equals zero. A nonoverlapping
condition associated with an edge-to-edge contact requires that the projection of the
two edges on the common perpendicular line of the two edges are
completely separated. The 6 possible edge-to-edge contacts lead to 3
independent conditions, i.e.,

\begin{equation}
\label{eq002} [{\bf v}_A-({\bf v}_B-{\bf
\boldsymbol{\lambda}}_1+{\bf \boldsymbol{\lambda}}_2+{\bf
\boldsymbol{\lambda}}_3)]\cdot{\bf l}_0 \ge 0 ,
\end{equation}

\begin{equation}
\label{eq003} [{\bf v}_C-({\bf v}_B-{\bf
\boldsymbol{\lambda}}_1+{\bf \boldsymbol{\lambda}}_3)]\cdot{\bf
l}_1 \ge 0 ,
\end{equation}

\begin{equation}
\label{eq004} [{\bf v}_C-({\bf v}_D+{\bf
\boldsymbol{\lambda}}_3)]\cdot{\bf l}_2 \ge 0 ,
\end{equation}
where ${\bf l}_0 = ({\bf v}_A-{\bf v}_E)\times({\bf v}_B-{\bf
v}_D)$, ${\bf l}_1 = ({\bf v}_B-{\bf v}_D)\times({\bf v}_E-{\bf
v}_C)$ and ${\bf l}_2 = ({\bf v}_D-{\bf v}_A)\times({\bf v}_C-{\bf
v}_E)$. Since all of the quantities involved in
Eqs.~(\ref{eq002})-(\ref{eq004}) are explicitly given, we do not
provide their lengthy forms here.

Furthermore, there are two additional nonoverlapping conditions
given by 4 potential vertex-to-face contacts, i.e.,

\begin{equation}
\label{eq005} {\bf \boldsymbol{\lambda}}_2 \cdot {\bf n}_1 \ge
\sqrt{\frac{2}{3}},
\end{equation}
where ${\bf n}_1 = (\frac{\sqrt6}{3}, -\frac{2\sqrt2}{6},
-\frac{1}{3})$ is the unit outward normal of the contacting face,

\begin{equation}
\label{eq006} ({\bf \boldsymbol{\lambda}}_1-{\bf
\boldsymbol{\lambda}}_2) \cdot {\bf n}_2 \ge \sqrt{\frac{2}{3}},
\end{equation}
and ${\bf n}_2 = (-\frac{\sqrt6}{3}, -\frac{2\sqrt2}{6},
\frac{1}{3})$ is the unit outward normal of the contacting face.
The edge-to-edge and vertex-to-face contacts are realized when the
equality holds in the above conditions
(\ref{eq5013})-(\ref{eq006}). Note we do not provide the lengthy
forms of the above equations because the quantities involved have
been given above. These nonoverlapping conditions constrain the
possible values of the remaining three variables.

Finally, the density of the six-parameter packings is given by
\begin{equation}
\label{eq50195} \phi =
\frac{4V_T}{V_F}=\frac{4V_T}{|{\boldsymbol{\lambda}}_1\times{\boldsymbol{\lambda}}_2\cdot{\boldsymbol{\lambda}}_3|},
\end{equation}

\noindent where $V_T = \sqrt2/12$ is the volume of a regular
tetrahedron with unit edge length and $V_F$ is the volume of the
fundamental cell given by

\begin{equation}
\begin{array}{c}
\label{eq50205} V_F = \frac{1}{3}(\alpha_2\beta_1+\alpha_1\beta_2)(-4\sqrt6+\sqrt6\alpha_1+3\sqrt2\alpha_2-2\alpha_3+2\sqrt6\beta_1+\beta_3) \\
-\frac{1}{12}(\alpha_3\beta_1+\alpha_1\beta_3)(-2\sqrt3\alpha_1-10\alpha_2+3\sqrt2\alpha_3+8\beta_2) \\
+\frac{1}{12}(-\alpha_3\beta_2+\alpha_2\beta_3)(-6\alpha_1-2\sqrt3\alpha_2+\sqrt6\alpha_3-2\sqrt6\beta_3).
\end{array}
\end{equation}

The packing density relation (\ref{eq50195})
in conjunction with (\ref{eq50205})
is a function of the 6 variables $\alpha_i$ and $\beta_i$ ($i~=~1,2,3$).
We optimize the density for this problem using a sequential linear programming method
subject to the nonoverlapping constraints given by Eqs.~(\ref{eq5013})-(\ref{eq006})
that specify all of the remaining possible contacts.
In this way, we obtain the densest known packings with density
$\phi = 0.856347\ldots$. This is a numerical verification of
the optimality of the these packings among all four-particle basis packings.
In the following section, we provide analytical constructions of the densest known
packings and other dense packings from our general six-parameter family.

\subsection{Three-Parameter Family of Dense Tetrahedron Packings}

Both our simulations and local numerical analysis of Eqs.~(\ref{eq50195}) and (\ref{eq50205})
suggest that realizing the face-to-face contacts associated with Eqs.~(\ref{eq5013}) and (\ref{eq5014})
and the edge-to-edge contact associated with Eq.~(\ref{eq002}) will lead to denser packings.
Therefore, we let the equality hold in Eqs.~(\ref{eq5013}), (\ref{eq5014}) and (\ref{eq002}). By
solving these three additional equations, we
can further eliminate three independent variables for the packing, i.e.,

\begin{equation}
\label{eq50155}
\begin{array}{c}
\alpha_2 = \sqrt3/2,\\
\end{array}
\end{equation}

\begin{equation}
\label{eq5015}
\begin{array}{c}
\displaystyle{\beta_2 =
\frac{1}{4}(-2\sqrt3\alpha_1+2\alpha_2+3\sqrt2\alpha_3)},\\
\end{array}
\end{equation}

\begin{equation}
\label{eq5016}
\begin{array}{c}
\displaystyle{\beta_3 =
\frac{1}{10}(-8\sqrt6+5\sqrt6\alpha_1+3\sqrt2\alpha_2+5\sqrt2\alpha_3+10\sqrt6\beta_1)}.
\end{array}
\end{equation}

Therefore, the lattice vectors and the centroids of the dimers are
functions of the three parameters $(\alpha_1, \alpha_3, \beta_1)$ only, i.e.,
\begin{equation}
\label{eq5018}
\begin{array}{c}
\displaystyle{{\bf c}_1 = (\frac{1}{5}-\frac{\alpha_3}{\sqrt6},~ -\frac{4}{5\sqrt3},~-\frac{\sqrt6}{5}-\alpha_3)}, \\\\
\displaystyle{{\boldsymbol{\lambda}}_1 = (-\alpha_1,~ -\frac{\sqrt3}{2},~ -\alpha_3)}, \\\\
\displaystyle{{\boldsymbol{\lambda}}_2 = (\beta_1,~ -\frac{\sqrt3}{4}+\frac{\sqrt3}{2} \alpha_1-\frac{3\sqrt2}{4}\alpha_3,~\frac{13\sqrt6}{20}-\frac{\sqrt6}{2}\alpha_1-\frac{1}{2}\alpha_3-\sqrt6 \beta_1)}, \\\\
\displaystyle{{\boldsymbol{\lambda}}_3 = (\frac{2}{5}-\alpha_1-\beta_1,~ -\frac{\sqrt3}{4}-\frac{\sqrt3}{2} \alpha_1+\frac{3\sqrt2}{4}\alpha_3,~ -\frac{21\sqrt6}{20}+\frac{\sqrt6}{2}\alpha_1-\frac{1}{2}\alpha_3+\sqrt6 \beta_1)}.\\
\end{array}
\end{equation}

The density of the packing is given by
\begin{equation}
\label{eq5019} \phi =
\frac{4V_T}{V_F}=\frac{\sqrt2}{3 V_F},
\end{equation}

\noindent where $V_F$ is the volume of the
fundamental cell given by
\begin{equation}
\label{eq5020} V_F = \frac{1}{100}(39\sqrt2 - 60\sqrt2 \alpha_1^2
+ 40\sqrt3 \alpha_1\alpha_3 +30\sqrt2 \alpha_3^2).
\end{equation}

\noindent Note that although the packing structures depend on all
three variables $(\alpha_1, \alpha_3, \beta_1)$, the density is
only dependent on two variables $(\alpha_1, \alpha_3)$. The feasible values of the
remaining variables are determined by
Eqs.~(\ref{eq002})-(\ref{eq006}). In particular, we have
\begin{equation}
\label{eq5021}
\begin{array}{c}
\displaystyle{\frac{3}{2}\alpha_1-\frac{5\sqrt6}{12}\alpha_3\le\frac{1}{20}},\\\\
\displaystyle{\frac{3}{2}\alpha_1-\frac{5\sqrt6}{12}\alpha_3\ge-\frac{1}{20}},\\\\
\displaystyle{max\{\frac{7}{10}-\alpha_1+\frac{\sqrt6}{6}\alpha_3,
~\frac{7}{10}-\frac{\sqrt6}{6}\alpha_3\}\le \beta_1 \le
min\{\frac{3}{4}+\frac{1}{2}\alpha_1-\frac{\sqrt6}{4}\alpha_3,
~\frac{3}{4}-\frac{3}{2}\alpha_1+\frac{\sqrt6}{4}\alpha_3\}}.\\
\end{array}
\end{equation}



\begin{figure}
\begin{center}
$\begin{array}{c}
\includegraphics[height=7.5cm, keepaspectratio]{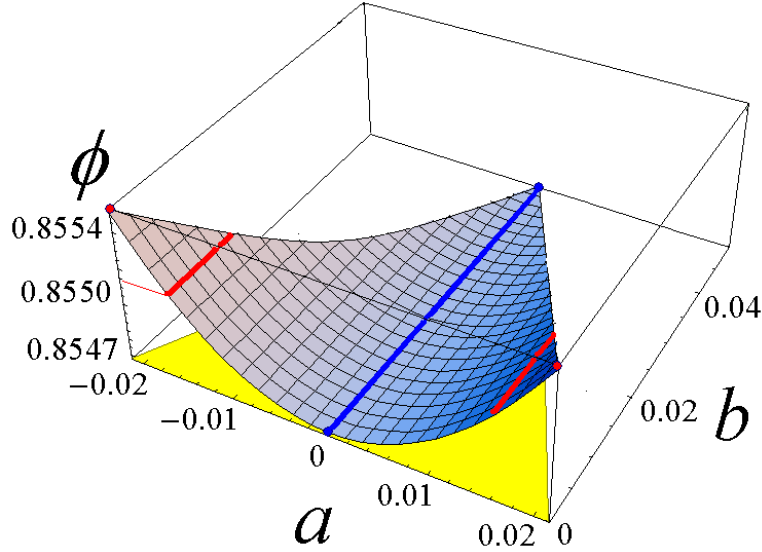} \\
\end{array}$
\end{center}
\caption{(color online). The density $\phi$ surface of our family
of tetrahedron packings as a function of the two parameters $a$
and $b$ (the parameters are related to the original variables via
$\alpha_1 =a$, $\alpha_3 = -\sqrt{\frac{2}{3}}a$, $\beta_1 =
\frac{3}{4}-2a-b$ for $\alpha_1>0$ and $\alpha_1 =a$, $\alpha_3=
-\sqrt{\frac{2}{3}}a$, $\beta_1 = \frac{3}{4}+a-b$ for
$\alpha_1<0$). As explained in the text, the thick red lines
(associated with $a\approx \pm 0.016$) show two sets of
tetrahedron packings with distinct structures but with the same
density. The two red points (associated with $a
=\pm\frac{3}{140}$) correspond to the densest two tetrahedron
packings in this family. The packings found by Kallus {\it et al.}
\cite{Ka09} are recovered from our two-parameter family (thick
blue line, associated with $a = 0$).} \label{fig1}
\end{figure}

From Eq.~(\ref{eq5019}), we see that maximizing the density is
equivalent to minimizing the volume of the fundamental cell $V_F$.
To obtain our two-parameter family of packings reported in Ref.~\cite{To10},
we assume that the minimum of $V_F$ can be obtained by optimizing
the two variables $(\alpha_1, \alpha_3)$ separately, i.e., we
assume the following is true:
\begin{equation}
\label{eq5022} min_{\{\forall(\alpha_1,\alpha_3)\}} V_F =
min_{\{\forall \alpha_1\}}[min_{\{\forall \alpha_3\}}V_F].
\end{equation}

\noindent The inner optimization (over $\alpha_3$ with fixed
$\alpha_1$) requires
\begin{equation}
\label{eq5023}
\partial V_F/\partial \alpha_3 = \frac{1}{5}(2\sqrt3 \alpha_1 + 3\sqrt2 \alpha_3) = 0
\end{equation}

\noindent which gives $\alpha_3 = -\sqrt{\frac{2}{3}}\alpha_1$.
This further reduces the number of variables and leads to a
two-parameter family of packings. Substituting this expression
into Eq.~(\ref{eq5020}), we obtain
\begin{equation}
\label{eq5024} V_F = \frac{1}{100}(39\sqrt2-80\sqrt2 \alpha_1^2).
\end{equation}

\noindent Substituting Eq.~(\ref{eq5024}) into Eq.~(\ref{eq5019})
leads to the density expression
\begin{equation}
\label{eq5025} \phi = \frac{100}{117-240\alpha_1^2}.
\end{equation}

\noindent where $\alpha_1 \in (-\frac{3}{140},\frac{3}{140})$. It
is important to note that for each $\alpha_1 \neq 0$, there are
two sets of packings of tetrahedra, each with distinct structures
but possessing the same density (as shown in Fig. \ref{fig1} by
the thick red lines). In addition, by substituting $\alpha_1 =a$,
$\alpha_3 = -\sqrt{\frac{2}{3}}a$, $\beta_1 = \frac{3}{4}-2a-b$
for $\alpha_1>0$ and $\alpha_1 =a$, $\alpha_3=
-\sqrt{\frac{2}{3}}a$, $\beta_1 = \frac{3}{4}+a-b$ for
$\alpha_1<0$, respectively into Eq.~(\ref{eq5018}), the two sets
of lattice vectors and dimer centroids can be obtained, i.e., for
$-\frac{3}{140}<a<0$, we have
\begin{equation}
\label{eq2}
\begin{array}{c}
\displaystyle{{\bf c}_1 = (\frac{1}{5}+\frac{a}{3},
-\frac{4}{5\sqrt3}, -\frac{3\sqrt2}{5\sqrt3}+
\sqrt{\frac{2}{3}}a)},\\
\displaystyle{{\bf \boldsymbol{\lambda}}_1 = (-a, -\frac{\sqrt3}{2}, \sqrt{\frac{2}{3}}a)}, \\
\displaystyle{{\bf \boldsymbol{\lambda}}_2 = (\frac{3}{4}+a-b, -\frac{\sqrt3}{4}+\sqrt3 a, -\frac{3}{5\sqrt6}-\frac{8}{\sqrt6}a+\sqrt6 b)}, \\
\displaystyle{{\bf \boldsymbol{\lambda}}_3 = (-\frac{7}{20}-2a+b,
-\frac{\sqrt3}{4}-\sqrt3 a,
-\frac{9}{5\sqrt6}+\frac{10}{\sqrt6}a-\sqrt6 b)},
\end{array}
\end{equation}
where $0<b<\frac{3+140a}{60}$; for $0<a<\frac{3}{140}$, we have
\begin{equation}
\label{eq3}
\begin{array}{c}
\displaystyle{{\bf c}_1 = (\frac{1}{5}+\frac{a}{3}, -\frac{4}{5\sqrt3}, -\frac{3\sqrt2}{5\sqrt3}+\sqrt{\frac{2}{3}}a)},\\
\displaystyle{{\bf \boldsymbol{\lambda}}_1 = (-a, -\frac{\sqrt3}{2}, \sqrt{\frac{2}{3}}a)}, \\
\displaystyle{{\bf \boldsymbol{\lambda}}_2 = (\frac{3}{4}-2a-b, -\frac{\sqrt3}{4}+\sqrt3 a, -\frac{3}{5\sqrt6}+\frac{10}{\sqrt6}a+\sqrt6 b)}, \\
\displaystyle{{\bf \boldsymbol{\lambda}}_3 = (-\frac{7}{20}+a+b,
-\frac{\sqrt3}{4}-\sqrt3 a,
-\frac{9}{5\sqrt6}-\frac{8}{\sqrt6}a-\sqrt6 b)},
\end{array}
\end{equation}
where $0<b<\frac{3-140a}{60}$.

\begin{figure}
\begin{center}
$\begin{array}{c@{\hspace{1.5cm}}c}
\includegraphics[height=5.0cm, keepaspectratio]{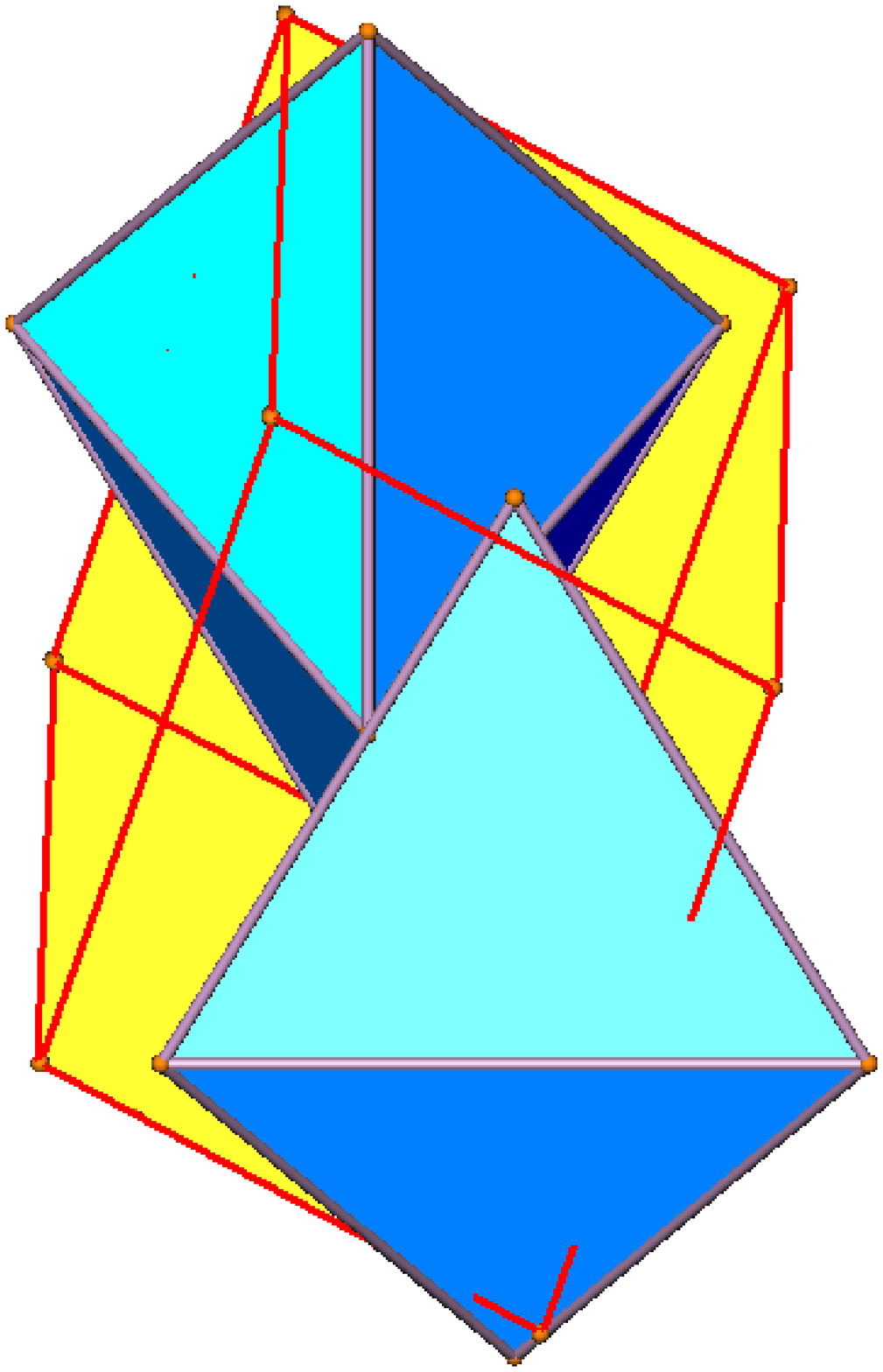} &
\includegraphics[height=5.0cm, keepaspectratio]{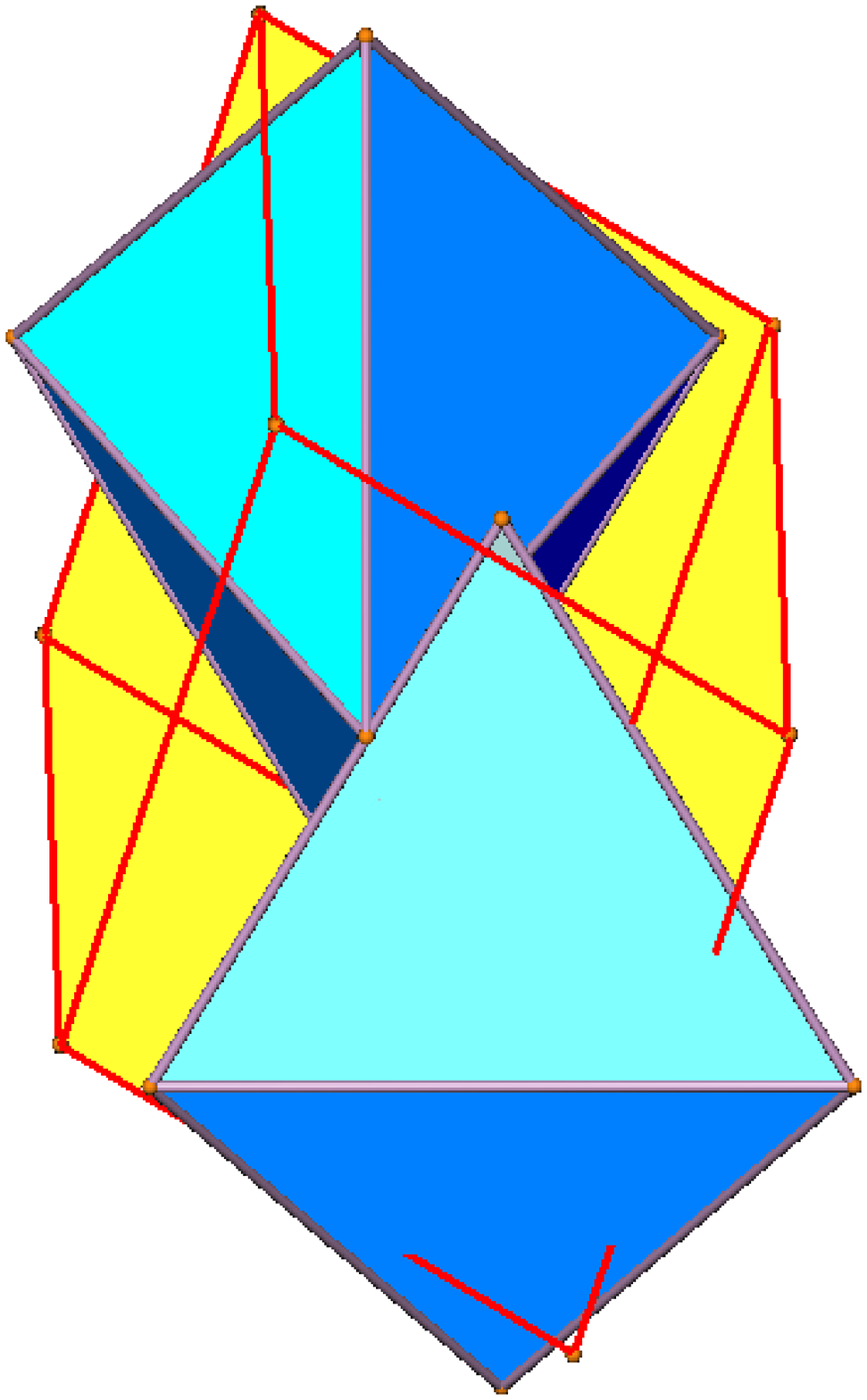} \\
\mbox{(a)} & \mbox{(b)}\\\\
\includegraphics[height=5.0cm, keepaspectratio]{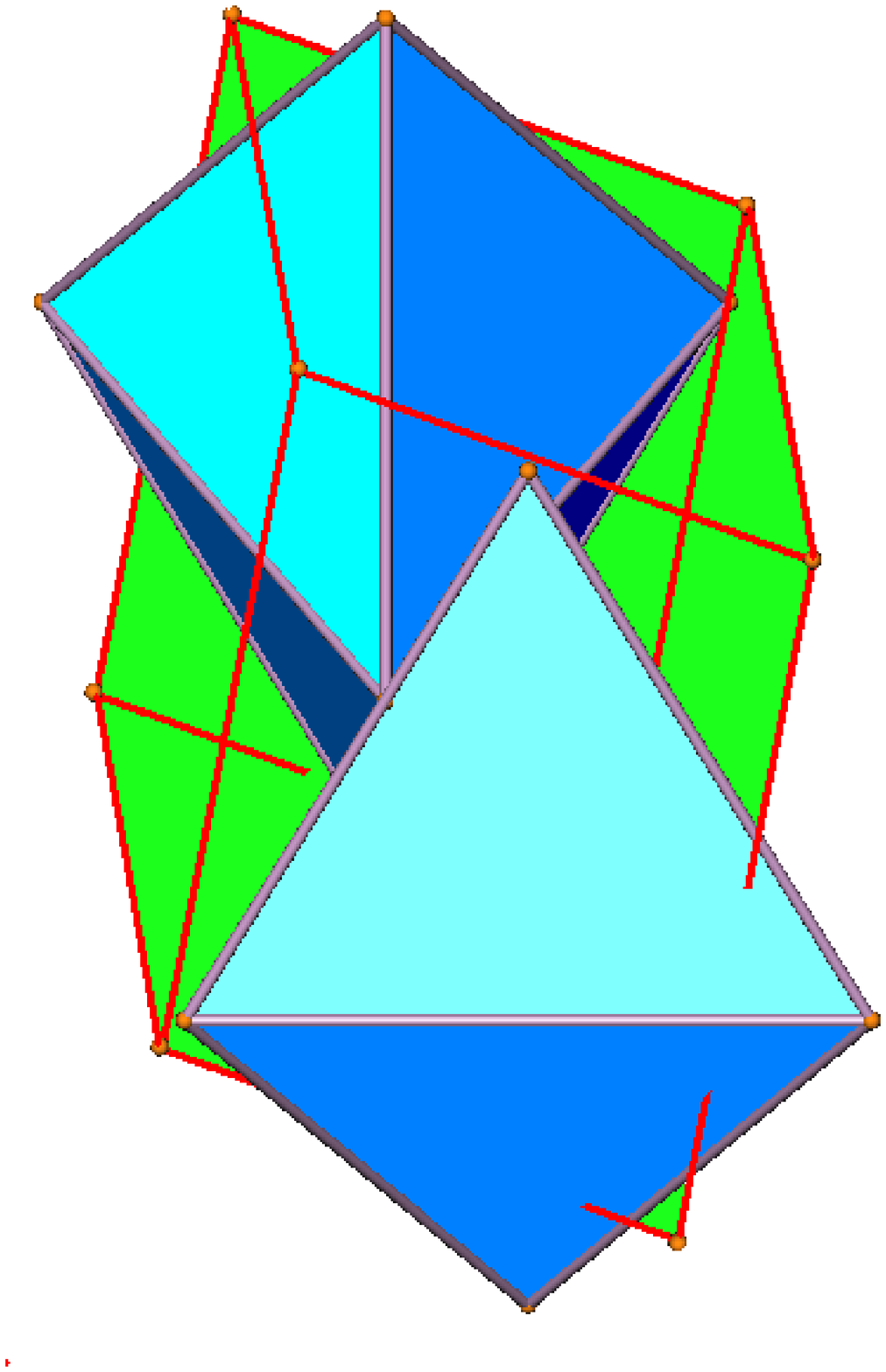} &
\includegraphics[height=5.0cm, keepaspectratio]{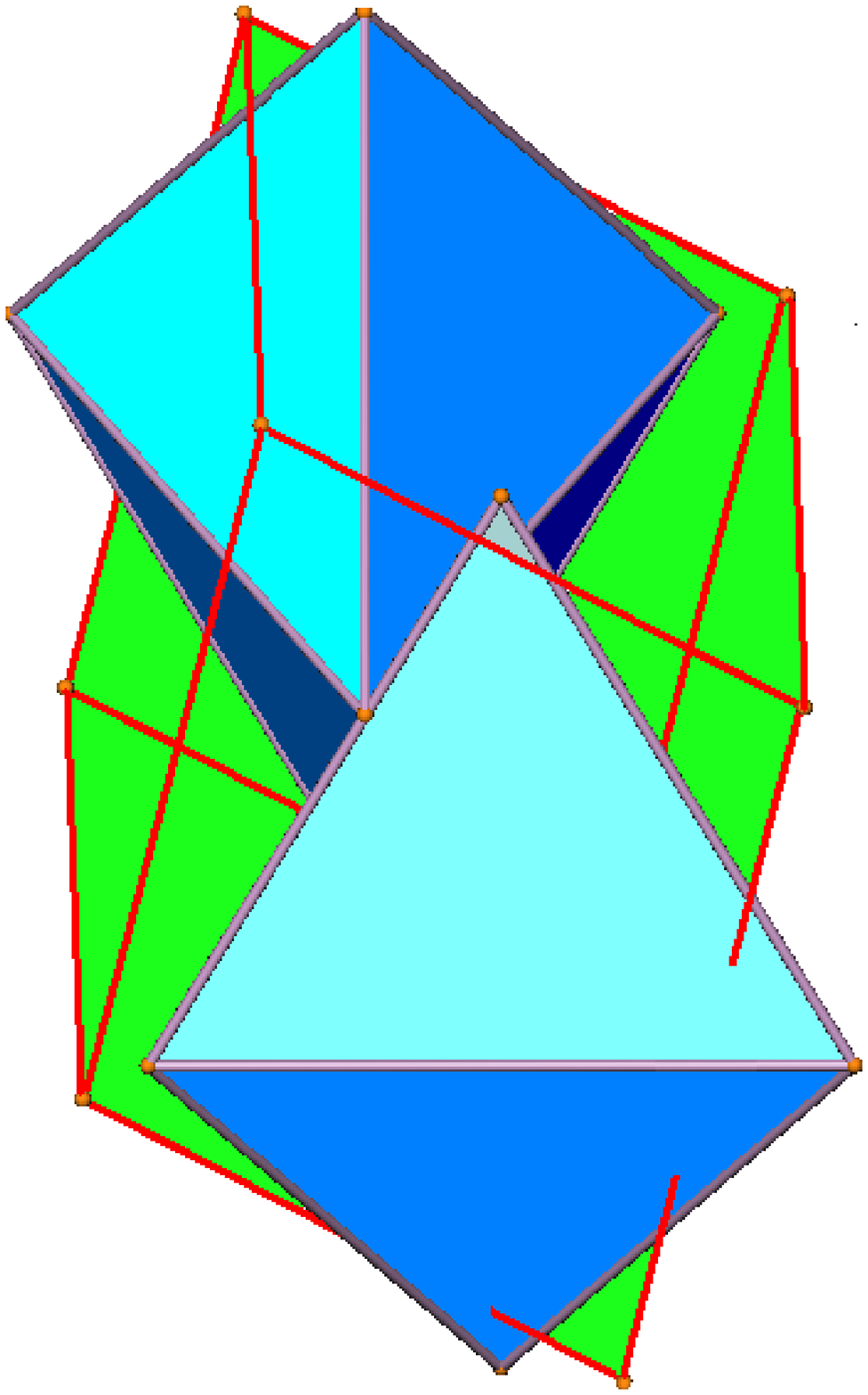} \\
\mbox{(c)} & \mbox{(d)}
\end{array}$
\end{center}
\caption{(color online). In the top panel [(a) and (b)], we show the two different
configurations of the densest packings of
four tetrahedra (two dimers) with $\phi = \frac{12250}{14319} = 0.855506\ldots$ within their corresponding
rhombohedral fundamental cells of the two-parameter family. (a)
Here $a=-\frac{3}{140}$ and $b=0$. (b) Here $a=\frac{3}{140}$ and $b=0$.
The fundamental cells are colored yellow (light gray) and their
boundaries are colored red (dark gray). The two packings and their
associated fundamental cells shown in (a) and (b) are only slightly
different from one another.
In the bottom panel [(c) and (d)], we show the two different
configurations of the densest packings of
four tetrahedra (two dimers) with $\phi = \frac{4000}{4671} = 0.856347\ldots$ within their corresponding
rhombohedral fundamental cells of the three-parameter family. (c)
Here $\alpha_1=\frac{7}{160}$, $\alpha_3 = \frac{3}{160}\sqrt{\frac{2}{3}}$
and $\beta_1=\frac{111}{160}$. (d) Here  $\alpha_1=-\frac{7}{160}$,
$\alpha_3 = -\frac{3}{160}\sqrt{\frac{2}{3}}$ and $\beta_1=\frac{59}{80}$.
The fundamental cells are colored green (medium gray) and their
boundaries are colored red (dark gray). The two packings and their
associated fundamental cells shown in (c) and (d) are only slightly
different from one another.} \label{fig2}
\end{figure}


\begin{figure}
\begin{center}
$\begin{array}{c@{\hspace{0.25cm}}c}
\includegraphics[height=5.25cm, keepaspectratio]{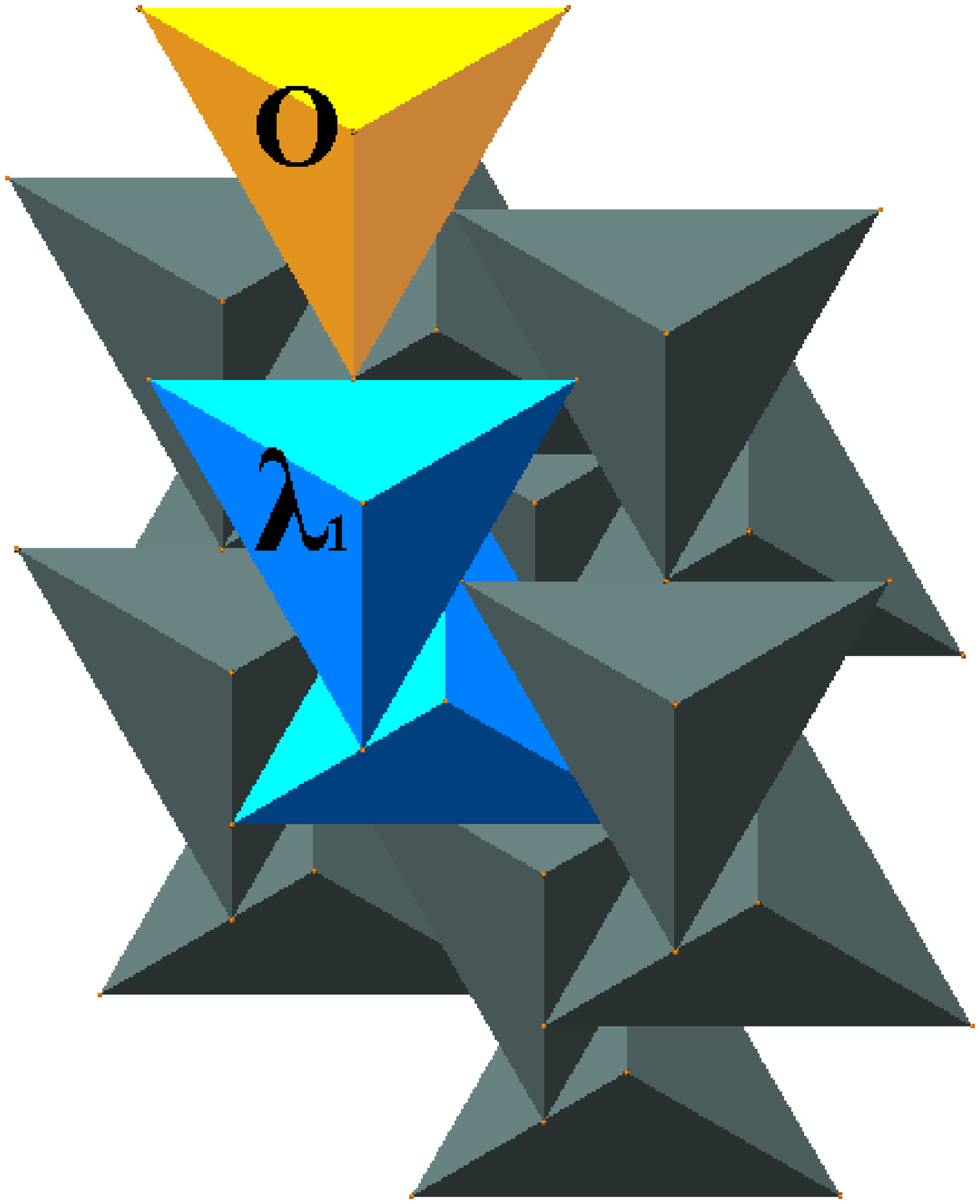} &
\includegraphics[height=5.25cm, keepaspectratio]{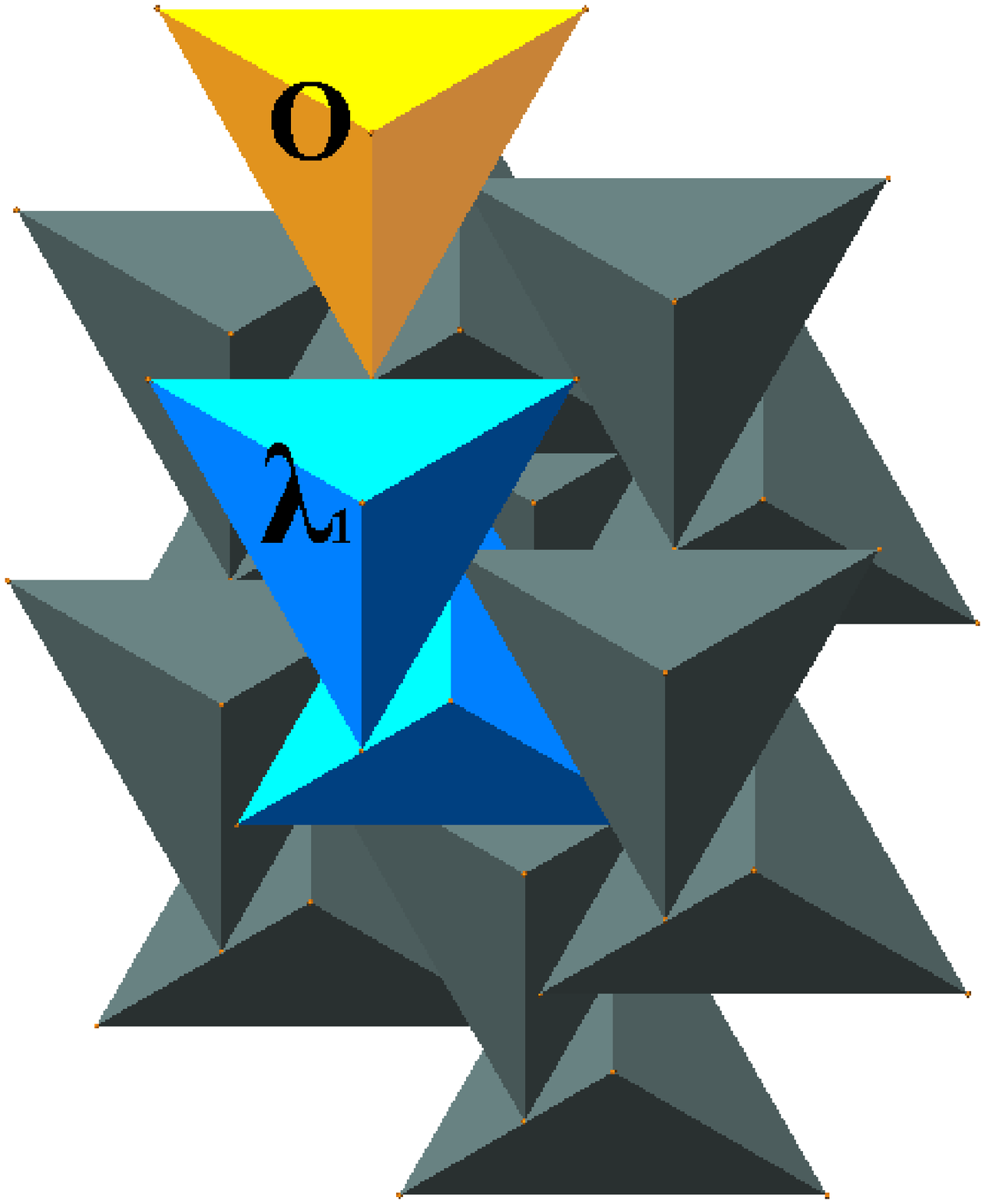} \\
\mbox{(a)} & \mbox{(b)}\\
\includegraphics[height=5.25cm, keepaspectratio]{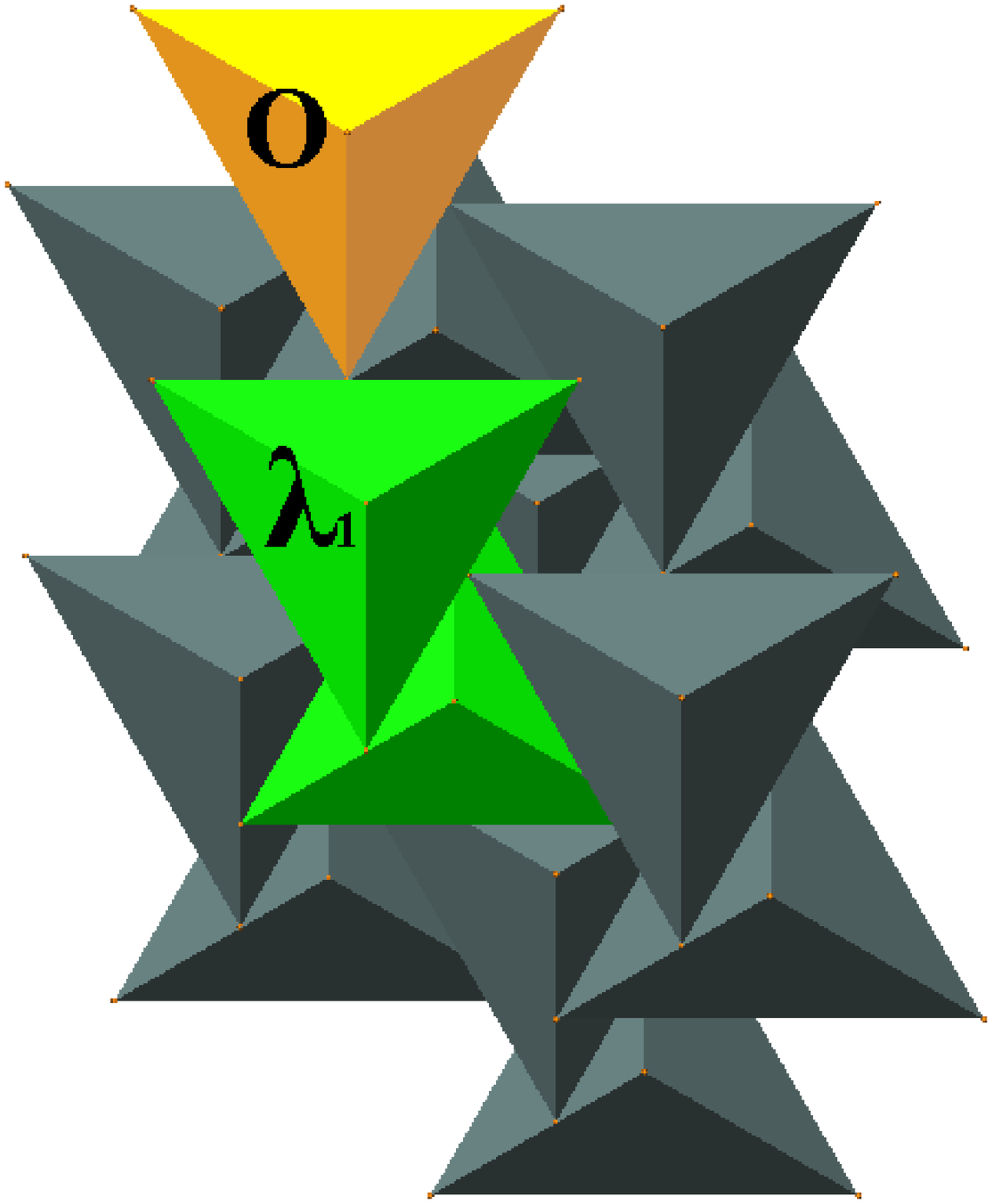} &
\includegraphics[height=5.25cm, keepaspectratio]{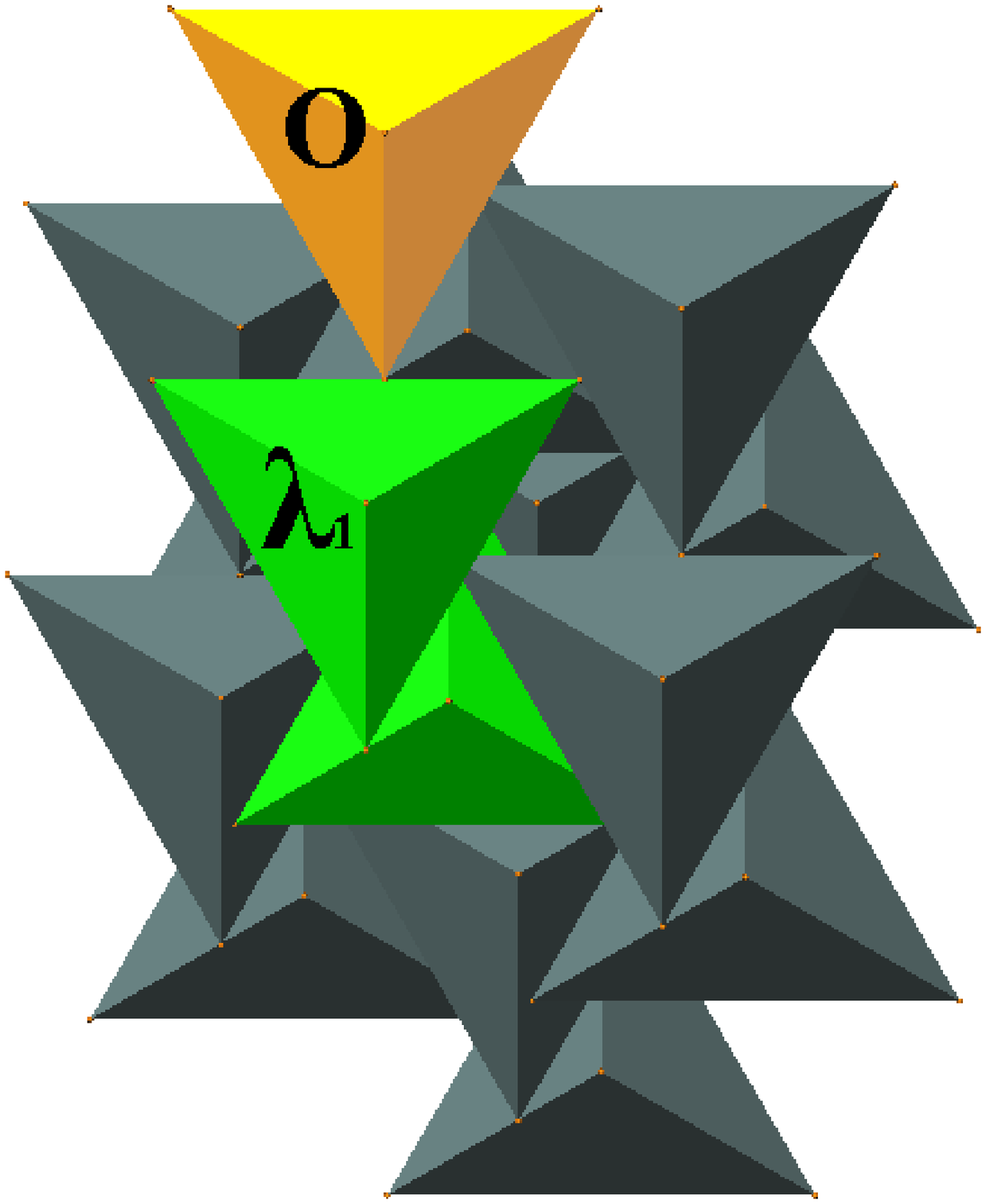} \\
\mbox{(c)} & \mbox{(d)}
\end{array}$
\end{center}
\caption{(color online). This figure shows periodic replicates the densest
tetrahedron packings of the two-parameter family (with $\phi = \frac{12250}{14319} = 0.855506\ldots$) and
the three-parameter family ( with $\phi = \frac{4000}{4671} = 0.856347\ldots$) corresponding to
the top and bottom panels in Fig. \ref{fig2}, respectively, with 8 fundamental cells (2 along each lattice
vector). The tetrahedra within the fundamental cells are shown in
blue and green (both appearing as medium gray in the print version) for the two-parameter [(a) and (b)] and
three-parameter [(c) and (d)] family packings, respectively.
Note that in (a), the dimer with centroid at ${\bf
\boldsymbol{\lambda}}_1$ (blue, or medium gray in the print version) is slightly shifted to the right
with respect to the dimer at the origin {\bf O} (yellow, or light gray in the print version); and in (b), the
dimer with the centroid at ${\bf \boldsymbol{\lambda}}_1$ (blue, or medium gray in the print version)
is slightly shifted to the left with respect to the dimer at the
origin {\bf O} (yellow, or light gray in the print version).
The three-parameter packing structures are very similar to that of the two-parameter packings.
Observe that in the perspective shown the
dimers appear as if they were single tetrahedra. } \label{fig3}
\end{figure}

The densest packings in this family are associated with
$a=-\frac{3}{140}$, $b=0$ and $a=\frac{3}{140}$, $b=0$, possessing
a density $\phi_{max} = \frac{12250}{14319} = 0.855506...$. In
each set, there is a unique packing structure associated with
$\phi_{max}$ (shown as the red points in Fig. \ref{fig1}), instead
of a spectrum of structures.
 Two different configurations of the densest packings
of the two-parameter family within their corresponding
rhombohedral fundamental cells are shown in Fig. \ref{fig2}(a) and (b).
The two four-particle configurations in the fundamental cells
are only slightly different from one another. Specifically, the difference
between the coordinates of the dimer centroids that are not at the
origin (i.e., at ${\bf c}_1$) is $(\frac{1}{70}, 0, \frac{\sqrt6}{70})$.
Figure \ref{fig3}(a) and (b) depict periodic replicates of the densest
tetrahedral packings corresponding to those shown in Fig.
\ref{fig2}(a) and (b) with 8 fundamental cells (2 along each lattice vector).
These packings are not chiral. The lattice vectors of packings
associated with positive and negative $a$ are related by an
isometric map \cite{Chen10}.

At $a=0$, there is only one set of packings with the same $\phi =
\frac{100}{117} = 0.85470...$, whose structures are dependent on $b$
(shown as the blue line in Fig. \ref{fig1}). These packings reduce
exactly to those discovered by Kallus et al., which possess
relatively high symmetry. In particular, $a=0$ allows the
centroids of the dimers, which are related to each other by an
integer multiple of ${\bf \boldsymbol{\lambda}}_1$,  to be
perfectly aligned on two of the mirror image planes of the dimers
simultaneously, which leads to additional two-fold rotational
symmetry of the packing. This additional rotational symmetry,
together with the point inversion symmetry, leads to uniform
packings with respect to each tetrahedron (not just each dimer),
i.e., the symmetry operation acts to take each tetrahedron into
another.


If we do not make the assumption used above (i.e.,
optimizing the two variables $\alpha_1$ and $\alpha_3$
separately), the boundary of the domain of $V_F$ needs to be
searched completely to find the minimal value. For the boundary
specified by
\begin{equation}
\frac{3}{2}\alpha_1-\frac{5\sqrt6}{12}\alpha_3\le\frac{1}{20},
\end{equation}
we have $\alpha_1=\frac{1}{90}(3+25\sqrt6 \alpha_3)$. Substituting
this expression into Eq.~(\ref{eq5020}) gives
\begin{equation}
\label{eq5026} V_F =
\frac{2}{1125}(219\sqrt2-5\sqrt3\alpha_3+200\sqrt2 \alpha_3^2)
\end{equation}
the minimal value of which is $V^*_F = \frac{1557}{2000\sqrt2}$
with $\alpha_3^* = \frac{3}{160}\sqrt{\frac{2}{3}}$. For the
boundary specified by
\begin{equation}
\frac{3}{2}\alpha_1-\frac{5\sqrt6}{12}\alpha_3\ge-\frac{1}{20},
\end{equation}
we have $\alpha_1=\frac{1}{90}(-3+25\sqrt6 \alpha_3)$.
Substituting this expression into Eq.~(\ref{eq5020}) gives
\begin{equation}
\label{eq5027} V_F =
\frac{2}{1125}(219\sqrt2+5\sqrt3\alpha_3+200\sqrt2 \alpha_3^2)
\end{equation}
the minimal value of which is $V^*_F = \frac{1557}{2000\sqrt2}$
with $\alpha_3^* = -\frac{3}{160}\sqrt{\frac{2}{3}}$.

Thus, we find a maximal density $\phi_{max} = \frac{4000}{4671}$,
which was also reported by Chen {\it et al.} \cite{Chen10},
associated with $\alpha_1 = \frac{7}{160}, \alpha_3 =
\frac{3}{160}\sqrt{\frac{2}{3}}$ and $\alpha_1 = -\frac{7}{160},
\alpha_3 = -\frac{3}{160}\sqrt{\frac{2}{3}}$. The packing
structures can be obtained by substituting  $\beta_1 =
\frac{3}{4}-\frac{3}{2}\alpha_1+\frac{\sqrt6}{4}\alpha_3 = \frac{111}{160}$ for
$\alpha_1>0$ and $\beta_1 =
\frac{3}{4}+\frac{1}{2}\alpha_1-\frac{\sqrt6}{4}\alpha_3 = \frac{59}{80}$ for
$\alpha_1<0$ into Eq.~(\ref{eq5018}). These three-parameter packings are very
close to the two-parameter packings in structure and possess the same symmetry
[see Fig.~\ref{fig2}(c) and (d), and Fig.~\ref{fig3}(c) and (d)].

\section{Towards Upper Bounds on the Maximal Density}
\label{bounds}

The problem of determining upper bounds on the maximal density of
packings of nonspherical particles is highly nontrivial, and yet
such estimates would be indispensable in assessing the packing
efficiency of a candidate dense packing, especially if tight upper
bounds could be constructed. It has recently been shown  that
$\phi_{max}$ of a packing of congruent nonspherical particles of
volume $v_{P}$ in $\mathbb{R}^3$ is bounded from above according
to
\begin{equation}
\phi_{max}\le  \mbox{min}\left[\frac{v_{P}}{v_{S}}\;
\frac{\pi}{\sqrt{18}},1\right], \label{bound}
\end{equation}
where $v_{S}$ is the volume of the largest sphere that can be
inscribed in the nonspherical particle and $\pi/\sqrt{18}$ is the
maximal sphere-packing density \cite{To09b,To09c}. The upper bound
(\ref{bound}) will be relatively tight for packings of
nonspherical particles provided that the {\it asphericity}
$\gamma$ (equal to the ratio of the circumradius to the inradius)
of the particle is not large. However, for tetrahedra, the
asphericity is too large for the upper bound (\ref{bound}) to
yield a result that is less than unity.

One possible approach to obtaining nontrivial upper bounds is to
attempt to generalize the idea that Rogers used to prove upper
bounds on $\phi_{max}$ for sphere packings \cite{Ro58}. The key
concept is to consider a locally dense cluster of 4 contacting
spheres in a tetrahedral arrangement and then prove that the
fraction of space covered by the spheres within the tetrahedron
joining the sphere centers is an upper bound on $\phi_{max}$. This
can be done because one can triangulate any sphere packing to
decompose it into generally irregular tetrahedra with vertices at
sphere centers. The fact that the regular tetrahedron has the best
density for any tetrahedron, then yields an upper bound for the
density of any sphere packing. In the case of the non-tiling
Platonic and Archimedean solids, a natural choice for the
enclosing region  associated with the cluster is its convex hull.

\begin{figure}
\begin{center}
$\begin{array}{c@{\hspace{0.35cm}}c}
\includegraphics[height=6.0cm, keepaspectratio]{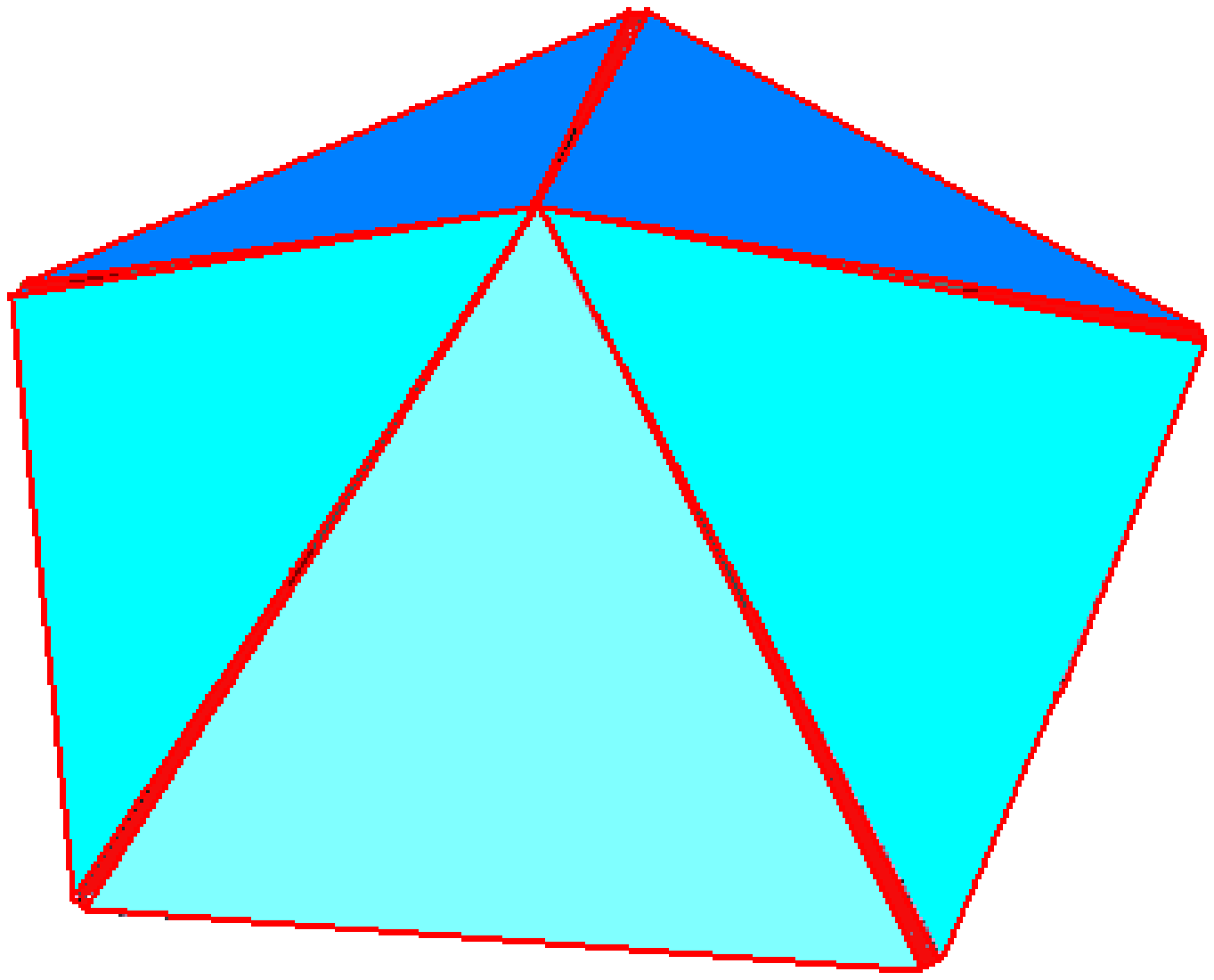} &
\includegraphics[height=5.5cm, keepaspectratio]{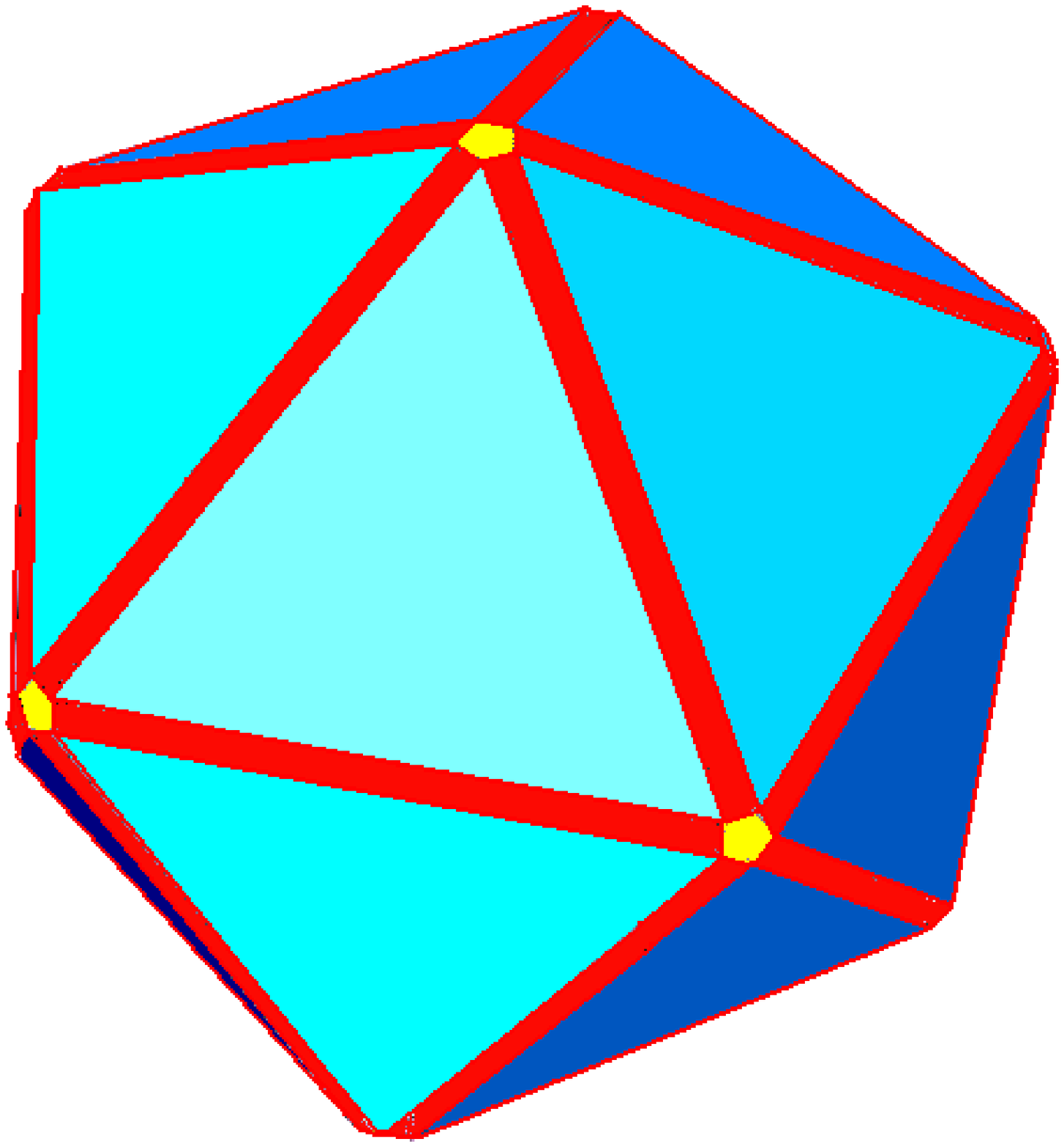} \\
\mbox{(a)} & \mbox{(b)}
\end{array}$
\end{center}
\caption{(color online). (a) The convex hull of five regular tetrahedra in a
``wagon wheel'' arrangement. The convex hull can be decomposed
into five regular tetrahedra (shown in blue, or medium gray in the print version) and five thin
irregular tetrahedra (shown in red, or dark gray in the print version). (b)The convex hull of 20
regular tetrahedra in an ``icosahedral" arrangement. The convex
hull can be decomposed into 20 regular tetrahedra (shown in blue, or medium gray in the print version),
12 pyramids with pentagonal bases (shown in yellow, or light gray in the print version) and 30 pyramid
with rectangular bases (shown in red, or dark gray in the print version).} \label{wagon}
\end{figure}

For tetrahedra, we must identify the densest local
cluster with density that exceeds or equals $\phi_{max}$ while taking into account
their shape symmetry. A trivial choice
is a dimer because the fraction of space covered by the dimer
within its convex hull is unity. Two nontrivial choices are the
5-particle ``wagon-wheel" cluster as shown in Fig. \ref{wagon}(a) and
a  more ``isotropic" icosahedron-like cluster (a cluster consisting of
20 tetrahedra that share a common vertex) as shown in Fig.~\ref{wagon}(b).
The ``wagon-wheel'' and icosahedron-like cluster can be considered to be the
most anisotropic and isotropic densest local clusters, respectively.
The former is essentially a ``flat" (quasi-two-dimensional) structure.
Since neither of the two local clusters can tile space, their
local packing densities, which are computed below,
provide limiting estimates on upper bounds for $\phi_{max}$.

The convex hull of the ``wagon-wheel'' cluster composed of five tetrahedra sharing a
common edge (a ``wagon wheel'' cluster) can be decomposed into
five regular tetrahedra and five thin irregular tetrahedra. We
assume the gaps between the regular tetrahedra are equal, i.e, the
thin irregular tetrahedra are congruent. Note the regular
tetrahedron shares two faces with its neighboring irregular
tetrahedra. Thus, the volume ratio is equal to the ratio of the
corresponding heights of the regular and irregular tetrahedron,
i.e.,
\begin{equation}
\displaystyle{\gamma = \frac{V_T}{V_{T_*}} =
\frac{1}{\frac{\sqrt2}{2}(3\cos^2\frac{3\pi}{10}-1)},}
\end{equation}
where $V_T$ and $V_{T_*}$ is the volume of the regular and
irregular tetrahedron, respectively. Thus, the density of this
local packing, defined as the fraction space covered by the
regular tetrahedra within the convex hull is given by
\begin{equation}
\displaystyle{\phi_W =
\frac{V_T}{V_T+V_{T_*}}=\frac{1}{\frac{3\sqrt2}{2}\cos^2\frac{3
\pi}{10}+(1-\frac{\sqrt2}{2})}=0.974857\ldots.}
\end{equation}
Because this cluster is highly anisotropic and is an effectively
``flat" object, it is reasonable to assume that it is not the
least densest local cluster and therefore its local density is likely to be
gross overestimate of the best upper bound on $\phi_{max}$.

It seems reasonable to suppose that the least densest local cluster
with the highest shape symmetry is one consisting of  20 tetrahedra sharing a common vertex,
i.e., an icosahedron-like cluster.
The convex hull of such a cluster can be decomposed
into 20 regular tetrahedra, 12 pyramids
with pentagonal bases and 30 pyramid with rectangular bases [see
Fig. \ref{wagon}(b)]. We assume the gaps between tetrahedra are
equal and thus the two types of pyramids are congruent. The volume
of the regular tetrahedron is $V_T = \frac{\sqrt2}{12}$ and the
volume $V_P$ of the pyramid with pentagonal base is given by

\begin{equation}
\displaystyle{V_P = \frac{5}{12}\tan(\frac{3\pi}{10}) L^2
\sqrt{1-\frac{L^2}{4 \cos^2\frac{3\pi}{10}}},}
\end{equation}
and the volume $V_R$ of the pyramid with rectangular base is given
by

\begin{equation}
\displaystyle{V_R = \frac{\sqrt2}{6}L\sqrt{1-L^2}},
\end{equation}
where

\begin{equation}
\displaystyle{L=(\frac{2\sqrt2}{\tau^2}-1)\sqrt{\frac{1}{6}+\frac{\sqrt5}{18}}},
\end{equation}
and $\tau = (1+\sqrt5)/2$ is the golden ratio. Thus, the local
packing density is given by

\begin{equation}
\displaystyle{\phi_I = \frac{20V_T}{20V_T+12V_P+30V_R} =
0.880755\ldots}.
\end{equation}

Note in the above calculations, we have assumed that the gaps
between the tetrahedra are equal, which is sufficient to provide
an estimate of the fraction of space covered by the cluster within
its convex hull.  It is noteworthy that the icosahedron-like
cluster of 20 tetrahedra is at least a local extremum, i.e.,
adding or removing a tetrahedron from the cluster will decrease
the local packing density. Moreover, adding another ``shell'' of
tetrahedra around the icosahedron-like cluster will also reduce
the local packing density. We note that the idea of Rogers to
prove an upper bound for sphere packings cannot be used here
because there is no analogous decomposition of space into
irregular convex hulls shown in Fig. \ref{wagon}(a) or (b). A
completely new idea is needed to prove that the aforementioned
estimates are bounds. If one could prove that these clusters are
indeed locally denser than the globally densest packings, then the
aforementioned estimates provide upper bounds on $\phi_{max}$, but
they cannot be sharp, i.e., they are not achievable by any
packings. Without loss of generality, it is reasonable to expect
that the globally densest packings of tetrahedra contain a
particular proportion of both the isotropic and anisotropic local
clusters. Therefore, these limiting estimates provide a plausible
range  of upper bounds, i.e., $\phi^{U}_{max}\in[0.880755,
~0.974857]$. It is noteworthy that the density $\phi
=0.856347\ldots$ of the densest known packings of tetrahedra is
relatively close to this putative lower-limit upper bound density
estimate of $0.880755\ldots$.

\section{Conclusions and Discussion}
\label{discuss}


For all of the small periodic packings  that we investigated
(including 2 to 32 particles per fundamental cell) using our
numerical ASC scheme, the densest packings that emerged had a
4-particle basis. The  numerical optimization  that we performed
of our six-parameter family of tetrahedron packings verified the
optimality of the densest 4-particle packing with density
$\phi=\frac{4000}{4671}=0.856347\ldots$. Are these dimer packings
optimal among all packings? This is a difficult question to answer
in any definitive way. We first note that the two contacting
dimers in the fundamental cell possess center inversion symmetry
and thus they can be viewed as a centrally symmetric compound
object. Hence, our two-parameter family of packings for the
two-dimers (and the densest known packings) are Bravais lattice
packings of such compound objects. Thus, it is not very surprising
that dense Bravais lattice packings of this centrally symmetric
compound object have a fairly high density based on the arguments
leading to Conjectures 1 and 2 \cite{To09b, To09c}, i.e., there
are a large number of face-to-face contacts which are made
possible due to the central symmetry of the compound object and
bring the centroids of the objects closer to each other. However,
Conjecture 3 states that central symmetry alone does not guarantee
optimality of the densest Bravais lattice packings among all
packings. That is, there are additional constraints on the
geometry and shape of the objects, e.g., they might also need to
possess three equivalent principal axes, as do the centrally
symmetric Platonic and Archimedean solids. When the latter
condition does not hold, there could exist non-Bravais lattice
packings that are denser than the optimal lattice packings (e.g.,
ellipsoids \cite{Do04d}). The 4-tetrahedron compound object is
very anisotropic and does not possess equivalent principal axes.
Moreover, it is a {\it concave} object rather than convex, which
makes it even more difficult to make a definitive conclusion about
the optimality of the densest dimer packings. In addition,
Haji-Akbari {\it et al.} \cite{Gl09} have found periodic packings
of tetrahedra consisting of 82 particles per fundamental cell with
a slightly lower density (i.e., $\phi \approx 0.8503$) than that
of the densest dimer packings. Their packings possess much larger
fundamental cells with distinctly different and more complicated
particle arrangements than the dimer packings. This suggests that
there is a large degeneracy of different  non-Bravais lattice
packing structures with high densities with various levels of
complexity and perhaps even higher densities than the optimal
4-particle packings. Such denser packings could be discovered by
carrying out exhaustive searches to determine the globally maximal
densities of packings with successively larger numbers of
particles per fundamental cell.


Previous numerical studies have indicated that dense packings may
have a large number of particles per fundamental cell arranged in
a complex fashion, e.g., the ``disordered wagon-wheels" packing
with $\phi = 0.822637\ldots$ \cite{To09c} and the ``ring stacks"
packing  with $\phi = 0.8503\ldots$ \cite{Gl09}. However, it is
now clear that such packings are in fact only locally optimal
solutions and hence the numerical techniques used to obtain them
are incapable of extricating themselves from these ``trapped"
regions of configuration space to find denser and more ordered
structures due to the intrinsic geometrical frustration of the
tetrahedron mentioned earlier. This may also call into question
claims made by Haji-Akbari {\it et al.} \cite{Gl09} that their
packings, which are characterized by an effective
``quasicrystal-like" plane, are true thermodynamic equilibrium
phases of tetrahedra at high densities. Instead, our
high-density constructions suggest that uniform periodic
packings with a 4-particle basis (or even some yet unknown denser
periodic packing) and their unjammmed, lower-density counterparts
could be the stable phases at high densities. If the latter is
correct and the putative ``quasicrystal-like" phase truly exists
at intermediate densities, then it is hard to imagine how such
complex quasicrystal-like structures of  tetrahedra under
quasi-static compression (densification) could rearrange
themselves at higher densities to spontaneously form a more
ordered periodic arrangement with higher symmetry. However, it is
difficult to draw any such definitive conclusions without further
study.

Although there could still be tetrahedron packings denser than our
constructions, it appears that all of the evidence thus far points
to the fact that the densest tetrahedron packings cannot possess
very high symmetry \cite{Co06,Ch08,To09b,To09c,Gl09} due to the
lack of central symmetry of a  tetrahedron and because tetrahedra
cannot tile space \cite{To09c}. Indeed, the dense 4-particle-basis
packings found in the present paper improved upon the best dimer
packings of Kallus {\it et al.} \cite{Ka09} by relaxing the
two-fold rotational symmetry constraints they imposed.


\begin{acknowledgments}
We are grateful to Henry Cohn, Yoav Kallus and John Conway for helpful discussions.
S. T. thanks the Institute for Advanced Study for
its hospitality during his stay there.
This work was supported by the Division of Mathematical Sciences
at the National Science Foundation under Award Number DMS-0804431
and by the MRSEC Program of the
National Science Foundation under Award Number DMR-0820341.
\end{acknowledgments}



\end{document}